\documentclass[a4paper]{article}

\usepackage[margin=1in]{geometry} 

\usepackage{amsmath}
\usepackage{amsthm}
\usepackage{amssymb}

\usepackage{booktabs}

\usepackage[utf8]{inputenc}
\usepackage{hyperref}
\hypersetup{
	unicode,
	pdfauthor={James Bono, Alec Xu},
	pdftitle={Randomized Controlled Trials for Security Copilot for IT Administrators},
	pdfproducer={LaTeX},
	pdfcreator={pdflatex}
}

\usepackage{natbib}
\setcitestyle{authoryear,open={(},close={)}} 

\theoremstyle{plain}

\theoremstyle{definition}

\usepackage{graphicx, color}
\graphicspath{{fig/}}
\usepackage{subcaption}
\usepackage{xcolor}

\usepackage{algorithm, algpseudocode} 
\usepackage{mathrsfs} 
\usepackage{multirow}

\usepackage{lipsum}

\title{Randomized Controlled Trials for Security Copilot for IT Administrators}
\author{James Bono\thanks{james.bono@microsoft.com} \and Alec Xu\thanks{alecxu@microsoft.com}}

\date{  
	Microsoft Corporation \\
	October 2024
}

\begin{document}
	\maketitle
	
	\begin{abstract}
    As generative AI (GAI) tools become increasingly integrated into workplace environments, it is essential to measure their impact on productivity across specific domains. This study evaluates the effects of Microsoft’s Security Copilot (``Copilot'') on information technology administrators (``IT admins'') through randomized controlled trials. Participants were divided into treatment and control groups, with the former granted access to Copilot within Microsoft's Entra and Intune admin centers. Across three IT admin scenarios -- sign-in troubleshooting, device policy management, and device troubleshooting -- Copilot users demonstrated substantial improvements in both accuracy and speed. Across all scenarios and tasks, Copilot subjects experienced a 34.53\% improvement in overall accuracy and a 29.79\% reduction in task completion time. We also find that the productivity benefits vary by task type, with more complex tasks showing greater improvement. In free response tasks, Copilot users identified 146.07\% more relevant facts and reduced task completion time by 61.14\%. Subject satisfaction with Copilot was high, with participants reporting reduced effort and a strong preference for using the tool in future tasks. These findings suggest that GAI tools like Copilot can significantly enhance the productivity and efficiency of IT admins, especially in scenarios requiring information synthesis and complex decision-making.
		
	\end{abstract}

	
\section{Introduction}
\label{sec:intro}
As generative AI (GAI) tools continue to mature, organizations are interested in understanding their impact on workforce productivity to guide adoption decisions. However, the effects of these tools can vary significantly depending on the specific domain and application. So, it is important to study each application individually. In this study, we evaluate the benefits of generative AI on information technology administrators (``IT admins'') by conducting a randomized controlled trial (RCT) with IT admins using Microsoft's Security Copilot (``Copilot''). We find that Copilot confers substantial benefits to IT admins, including accuracy improvements across all scenarios and tasks (34.53\%*** on average) and time savings across all scenarios (29.79\%*** on average).\footnote{Throughout, we indicate statistical significance using asterisks, where * indicates $0.05<p\leq 0.1$, ** indicates $0.01<p\leq 0.05$, and *** indicates $0<p\leq0.01$.}

As an add-on feature of Microsoft's identity management (Entra) and device management (Intune) solutions, Copilot's core functionality consists of its ability to reason across the vast data these solutions contain. In identity management, this includes all the existing identities, sign-in logs, conditional access policies, alerts about risky behavior, etc. In device management, this includes all the existing devices, policies, compliance states, etc. When configuring policies and troubleshooting devices and sign-ins, Copilot can extract and summarize salient information from multiple sources, giving IT admins the complete context without needing to compile the information themselves. 

We recruited 182 subjects and asked them to sign into an environment created for this experiment. Half of the subjects were assigned to the control group and were granted access to the Entra and Intune admin centers. The other half were assigned to the treatment group and were additionally given access to Copilot within the Entra and Intune admin centers. Then, subjects used these tools and the simulated data in that environment to respond to our experimental tasks. 

Our experimental tasks correspond to three scenarios in the identity management and device management domains. In the identity management domain, we study sign-in troubleshooting with tasks that mimic the kinds of activities that IT admins regularly engage in when diagnosing sign-in issues, including information retrieval and logs understanding. In the device management domain, we study both policy management and device troubleshooting. The policy management tasks focus on understanding the effects of policy settings on users and security. The device troubleshooting tasks require subjects to diagnose device non-compliance and which policies could be responsible for user-reported device issues. Our study includes multiple choice (MC), select-all-that-apply (SATA), and free response tasks. For each task, we measure completion time and accuracy. We also measure overall user satisfaction.

The following is a high-level summary of our primary research questions and answers:
\begin{itemize}
\item \textbf{RQ1: Does Copilot offer productivity improvements in common IT admin scenarios?} 
We find statistically significant improvements in speed and accuracy for all three scenarios. \newline

\item \textbf{RQ2: Do Copilot's productivity effects differ by subjects' experience levels in device management and identity management?}
We estimate greater effects for less experienced subjects. However, the differences are not statistically significant. \newline
    
\item \textbf{RQ3: How do Copilot's productivity effects differ according to task type, i.e., free response versus MC and SATA?} The benefits in speed and accuracy increase in the complexity of the task, with free response tasks garnering the greatest improvements. MC tasks were the least affected.\newline 

\item \textbf{RQ4: Are subjects satisfied with Copilot's capabilities and user experience?} Copilot subjects found the task less draining and requiring less effort than the control group, and they correctly perceived speed and accuracy improvements. They also indicated they would like to use Copilot the next time they do such tasks. 
\end{itemize}

This paper is organized as follows. First, we review related work. Next, we provide a detailed description of our experimental methods. Then, we present the results of our analysis and address the research questions above. Finally, we discuss the implications of these findings and future work.

\section{Related Work}
\label{sec:related}
The question of GAI's impact on productivity is central in driving both firm-level adoption decisions and economic policy decisions. Studies that examine GAI's productivity effects from a macroeconomic perspective often arrive at different conclusions about their likely magnitude. Looking at global patent and publication data, \citet{Parteka2023} fail to find a strong relationship between GAI and firm productivity. Projecting into the future, \citet{Acemoglu2024} similarly suggests the effects will be modest, no more than a 0.66\% increase in total factor productivity over 10 years. Other studies suggest the opposite. Looking at survey data on GAI adoption for a sample of German firms and controlling for endogeneity, \citet{Czarnitzki2023} find robust positive and significant associations between GAI adoption and firm productivity. \citet{Gao2023} look at micro-level manufacturing data and estimate that every 1\% of GAI penetration leads to 14\% increase in total factor productivity. On uptake, \citet{Bick2024} conduct a survey of US workers and find that GAI adoption has been faster than the personal computer and the internet with one in nine workers using it every workday. 

To complement the macroeconomic perspective, many studies have turned to measuring domain and task-specific productivity effects through laboratory or field experiments. Our work adds to this growing literature exploring GAI's potential to improve performance across various domains and task types. The overarching theme of this literature is to provide insights into the specific conditions under which GAI delivers the most value. Like our study, most of this work shows that GAI offers significant productivity benefits to skilled workers. In the software development domain, \citet{Peng2023} conducted a controlled experiment that asked developers to complete programming tasks with and without the assistance of GitHub Copilot. The results showed a 55.8\% reduction in task completion time for the group using Copilot. Similarly, \citet{Noy2023} observed substantial productivity improvements in professional writing tasks through the use of ChatGPT. In their experiment, GAI reduced task completion times by 40\% and increased output quality by 18\%. Other laboratory studies have shown similar gains from GAI tools across a variety of domains (\citet{Edelman2024, Freeman2024, Choi2023, DellAcqua2023}). 

Results from recent field experiments have also provided evidence of GAI's productivity benefits. \citet{Brynjolfsson2023} conducted a field experiment with over 5,000 customer support agents and demonstrated that access to a GAI-based conversational assistant improved productivity by 14\%, measured as issues resolved per hour. This parallels findings from \citet{Cui2024}, who conducted three field experiments with software developers using GitHub Copilot and reported a 26.08\% increase in tasks completed. Given the commonalities between IT admin tasks and the domains that have been studied in previous experiments -- software development, security analysis, consulting, legal analysis, and customer support -- the similar effects we measure are, to some degree, expected. 

There is also mounting evidence that GAI offers greater productivity benefits to lower-skilled workers (\citet{Peng2023, Cui2024, Choi2023}). The \citet{Brynjolfsson2023} study revealed that novice customer support agents benefited the most, showing a 34\% improvement compared to minimal improvements for highly skilled workers. \citet{Edelman2024} also found different effects for novices and professionals. In security incident investigation tasks, novices experienced a 35\% quality improvement with Copilot versus just a 7\% improvement for professionals. And across a set of 18 consulting tasks, \citet{DellAcqua2023} found that lower-skilled consultants improved by 43\% with GAI compared to just 17\% for higher-skilled consultants. 

Finally, there is evidence across the literature that the benefits of GAI are task-specific (\citet{DellAcqua2023}). This is consistent with theoretical predictions about the impact of GAI on labor markets (\citet{Agrawal2019,Frank2019,Eloundou2023}). It is also encoded in the approach taken by \citet{Felten2023} that characterizes occupational exposure to GAI by categorizing the tasks required by each occupation and thus its overlap with the things GAI can do well. On task heterogeneity, we find that GAI provides greatest productivity improvements in our free response tasks, which require open-ended reasoning over multiple data sources, compared to simpler information-retrieval tasks. This observation helps our understanding of how organizations might deploy GAI tools like Copilot effectively -- targeting areas where GAI support can enhance information synthesis and decision-making.
	
\section{Methodology}
\label{sec:methods}
We follow a standard RCT protocol to identify and estimate the causal effect of Copilot on accuracy and task completion time. That is, we randomly assign half of our subjects to the treatment group, which gives them the ability to use the Entra and Intune admin centers \textit{with} Copilot in responding to the experimental tasks. The other half of the subjects are assigned to the control group and have use of the Entra and Intune admin centers \textit{without} Copilot. To make our design as realistic as possible, we placed no restrictions on how subjects used other tools, such as other AI tools or web searches.\footnote{Indeed, after they completed the task, we asked control subjects whether they used any other AI tools, and 25\% indicated they used some AI tool to help them in this task -- most said they used ChatGPT.} Therefore, our measurements reflect the incremental benefit of Copilot on top of the admin centers and other tools like web searches and non-Copilot AI. This makes our estimates more conservative than they would be if we had tightly restricted tool use beyond Copilot.

We recruited our subject pool through Upwork, a marketplace for freelancers. We required subjects to be proficient in reading and writing English and to have a positive reputation on Upwork. We told them to expect the tasks to take approximately two hours to complete. We offered performance incentives for combined speed and accuracy. Specifically, we multiplied the percentiles of each subject's speed and accuracy and awarded payments on the following scale:
\begin{itemize}
    \item Top 10\% in combined speed and accuracy: \$130 
    \item Top 10-20\% in combined speed and accuracy: \$110 
    \item Top 20-30\% in combined speed and accuracy: \$90 
    \item Completed in good faith but did not reach the top 30\% in combined speed and accuracy: \$70 
    \item Show up fee: \$20
\end{itemize}

We had 181 subjects complete the task satisfactorily.\footnote{We dropped one subject who finished in unreasonably short time and scored worse than guessing percentage.} Our subject pool embodied a diverse range of experience levels across identity and network device management. See table \ref{tab:subjects} for a full breakdown of subjects by experience level. 

\begin{table}[ht]

\centering
\begin{tabular}{l c c c c c}
\toprule
\textbf{Device Management} & \multicolumn{4}{c}{\textbf{Identity Management Experience}} & \\
\cmidrule(lr){2-5}
\textbf{Experience} &  \textbf{0-1 years} & \textbf{1-3 years} & \textbf{3-5 years} & \textbf{5+ years} & \textbf{Total} \\ 
\midrule 
\textbf{0-1 years}          & 49 & 26 & 12 & 6  & 93  \\ 
\textbf{1-3 years}          & 8  & 19 & 13 & 5  & 45  \\ 
\textbf{3-5 years}          & 3  & 2  & 12 & 6  & 23  \\ 
\textbf{5+ years}         &  1  & 4  & 4  & 11 & 20  \\ 
\midrule
\textbf{Total} & \textbf{61} & \textbf{51} & \textbf{41} & \textbf{28} & \textbf{181} \\
\bottomrule
\end{tabular}
\caption{Subject counts by years of experience in network device management and identity management.\label{tab:subjects}}
\end{table}

We built our laboratory within the cloud-based identity and network device admin centers, respectively Entra and Intune. We simulated data in these environments to reflect what might be found in the environment of a small organization. The data involved a variety of identities, devices, sign-in logs, policies, and risk and compliance states. Subjects were given unique identities with read access to the environment and were asked to log in. They were then given instructions on how to navigate the admin centers to find the pages with information relevant to the experimental tasks.\footnote{See the appendix for subject feedback on instruction clarity.} The treatment subjects were additionally given instructions on how to use Copilot in the experimental tasks.

We presented subjects with experimental tasks across three IT admin scenarios. We chose these scenarios for their central importance in IT admin work, and we designed the tasks to resemble the kind of information retrieval and decision-making commonly required in the scenarios.
\begin{enumerate}
    \item \textbf{Sign-in Troubleshooting (Entra):}
    Diagnosing sign-in issues and investigating access anomalies are critical tasks for IT admins. This process typically involves reviewing detailed sign-in logs, identifying patterns, and cross-referencing data to uncover potential causes of failure. This scenario addresses a core responsibility where timely, accurate information is essential to maintaining security and access continuity. The experimental tasks require retrieving sign-in log information and analyzing it for potential issues, which aligns directly with the daily demands of IT admins, making this scenario highly relevant.
    \begin{enumerate}
        \item \textit{(SATA) Which users signed in with compliant devices on a specific date?} Subjects were presented with a list of seven users of which two signed in with compliant devices.
        \item \textit{(Free Response) Which conditional access policies were responsible for sign-in failures on a specific date?} We asked for responses in simple list format that included the conditional-access policies that failed and the users and applications affected by each. There were three conditional-access policies that failed across three users and one application.
        \item \textit{(Free Response) Summarize a user's sign-ins from a specific date.} We asked subjects to aggregate key details across sign-ins in a way that would be useful for an IT admin troubleshooting the user's sign-in issues.
    \end{enumerate}
    
    \item \textbf{Device Policy Management (Intune):}
    Managing and configuring device policies is a routine but complex task that impacts both security and usability within an organization. These policies influence how devices comply with organizational standards and affect users directly. The policy management tasks in this scenario focus on understanding the implications of specific policy settings—skills that are crucial for IT admins tasked with ensuring device compliance while balancing security and user needs. These tasks reflect the strategic thinking admins need to manage policies effectively, making them directly relevant to practical device management.
    \begin{enumerate}
        \item \textit{(Free Response) Summarize the potential impact of a given policy on users.} We asked subjects to highlight the key settings that will impact users, the impacts they will have, and why.
        \item \textit{(MC) We asked four questions about the effects of specific policy settings on users and security.}
    \end{enumerate}
    
    \item \textbf{Device Troubleshooting (Intune):}
    Diagnosing and resolving device compliance issues is a time-sensitive task for IT admins, as it directly impacts both user productivity and organizational security. The troubleshooting tasks in this scenario, such as identifying non-compliance causes and pinpointing problematic policies, mirror the common challenges admins face when resolving device issues. The relevance of this scenario is grounded in the fact that IT admins must quickly and accurately diagnose these problems to maintain a secure and operational IT environment.
    \begin{enumerate}
        \item \textit{(SATA) For which of the following settings is the given device noncompliant?} There were six options, and three were correct.
        \item \textit{(SATA) We asked two questions requiring subjects to identify which policies may be causing an issue with a device.}
    \end{enumerate}
    
\end{enumerate}

There are important differences between MC, SATA, and free response tasks, particularly in complexity, which we define based on the cognitive demands and effort required to complete the tasks. MC tasks represent the lowest level of complexity. They require subjects to select one correct answer from a predefined list of options, which provides clear boundaries, enabling subjects to reduce search and cognitive effort. SATA tasks are moderately complex. Unlike MC tasks, subjects must evaluate each option independently, since there are potentially multiple correct answers, which requires more time and consideration of a broader set of information sources. Finally, free response tasks are the most complex. Subjects must generate their own responses rather than select from a predefined list of options. Our free response tasks require subjects to focus their efforts on the information that matters and to synthesize information from multiple sources within the admin centers. We include all three task types in our experiments in order to see how Copilot's effect varies across them. 

Each task was graded on a scale from zero to one. For MC tasks, we awarded one for correct responses and zero for incorrect responses. For SATA tasks, we treated each option as a separate binary (true/false) question and scores are proportional to the number of options the subject classified correctly.\footnote{Our results are robust to alternative grading schemes.} For free response tasks, we used an LLM-based grader to identify whether the response contained each of a set of key facts, where the final score is the proportion of key facts included. This accuracy grade does not consider the quality of the writing. We assign equal weight to all questions in reporting aggregate accuracy scores. The maximum possible accuracy score is 11 points. In addition to accuracy, we prompted the LLM grader to judge the quality of free responses in terms of their clarity and organization. These ``quality'' scores are indicated separately. 

It is uninformative to compare speeds across groups when one group is systematically more accurate than another, which, as discussed in the next section, is the case here. In fact, to optimize their expected payoff under our incentive structure, which rewards both speed and accuracy, a rational subject might quickly assess that they are better off skipping a question that they have low expectation of answering correctly in reasonable time. Therefore, we use statistical methods to hold accuracy constant when examining Copilot's effect on task completion time. This leads us to our examination of the time that \textit{would be required} to achieve comparable accuracy. The details of this methodology are shared in the appendix (\citet{Edelman2024}). 
	
\section{Results}
\label{sec:results}
The results answer RQ1 in the affirmative: Copilot \textit{does} offer speed and accuracy improvements in common IT scenarios. Across all tasks, we measure a 34.53\%*** increase in accuracy and a 29.79\%*** reduction in task completion time. In the subsections on accuracy and speed below, we present detailed results answering RQ1-RQ4.

\subsection{Accuracy}
\label{sec:accuracy}
We measure statistically significant accuracy improvements in all scenarios. Table \ref{tab:accuracy} summarizes these results by task type and scenario. The ``Aggregate Results, All Content'' numbers represent our headline accuracy effect, a 34.53\%*** improvement for Copilot users. It includes the accuracy scores for all scenarios and question types, but it does not include quality scores for free response questions. 

\begin{table}[ht]
\centering
\begin{tabular}{l c c c c}
\toprule
& \textbf{Possible Points} & \textbf{Treatment} & \textbf{Control} & \textbf{Improvement} \\ 
\midrule
\multicolumn{5}{l}{\textbf{Aggregate Results}} \\ 
All Content & 11 & 6.97 & 5.18 & 34.53\%*** \\ 
SATA & 4 & 3.30 & 2.71 & 21.76\%*** \\
MC & 4 & 2.01 & 1.80 & 11.99\% \\
MC \textit{(Points $\geq$ 1)} & 4 & 2.09 & 1.975 & 11.39\%* \\ 
Free Response Content & 3 & 1.66 & 0.67 & 146.07\%*** \\ 
\midrule
\multicolumn{5}{l}{\textbf{Sign-In Troubleshooting}} \\ 
All Content & 3 & 1.80 & 1.23 & 46.88\%*** \\ 
SATA & 1 & 0.84 & 0.74 & 13.13\%** \\ 
Free Response Content & 2 & 0.97 & 0.49 & 97.75\%*** \\ 
\midrule
\multicolumn{5}{l}{\textbf{Device Policy Management}} \\ 
All Content & 5 & 2.70 & 1.98 & 36.39\%*** \\ 
MC & 4 & 2.01 & 1.80 & 11.99\% \\
MC \textit{(Points $\geq$ 1)} & 4 & 2.09 & 1.975 & 11.39\%* \\ 
Free Response Content & 1 & 0.69 & 0.18 & 275.15\%*** \\ 
\midrule
\multicolumn{5}{l}{\textbf{Device Troubleshooting}} \\ 
SATA & 3 & 2.47 & 1.97 & 24.99\%*** \\ 
\bottomrule
\end{tabular}
\caption{Copilot's effect on accuracy by question type and scenario. The top section shows aggregate results, while the bottom three sections compare task scores by individual scenarios.}
\label{tab:accuracy}
\end{table}

Copilot shows statistically significant effects across all three scenarios. In the identity management domain, our sign-in troubleshooting scenario yielded an ``All Content" accuracy improvement of 46.88\%***. In the device management domain, the policy management scenario yielded an accuracy improvement of 36.39\%***, and the device troubleshooting scenario yielded an accuracy improvement of 24.99\%***. 

Copilot shows statistically significant accuracy improvements in two out of three task types. The effect sizes generally track the task complexity, with the greatest benefits arising in free response tasks where Copilot subjects reported 146.07\%*** more relevant facts than our control group. In contrast to MC and SATA tasks, which only require selecting the correct answers from a short list of possibilities, the free response tasks leverage AI's unique ability to find and report relevant facts about a given scenario from a large and complex corpus, e.g., looking through sign-in logs to find and categorize sign-in failures caused by conditional access policies. Overall, Copilot yields an accuracy improvement of 21.76\%*** on SATA tasks across the sign-in troubleshooting and device troubleshooting scenarios. 

Although Copilot's estimated effect on accuracy for the device policy management MC tasks is positive, 11.99\%, this effect is the only one lacking statistical significance. There are two reasons. The first is that MC tasks take less advantage of the capabilities of GAI than the other question types, thus suggesting a smaller anticipated effect size. The second reason has nothing to do with technology but is rather an artifact of the all-or-nothing nature of the MC scoring, which necessarily degrades statistical power for a given sample size. To see this, we compute the coefficient of variation for the control group in each scenario and task type. Here, higher numbers reflect lower statistical power regardless of effect size. We get 34.6\% for the sign-in troubleshooting SATA scores, 43.1\% for the device troubleshooting SATA scores, and 56.1\% for the device policy management MC scores. Hence, the lack of statistically significant effects in the device policy management MC task perhaps says more about the task type than it does about Copilot's effectiveness in device policy management. Indeed, the free response tasks in the same device policy management scenario tested similar knowledge. On these, Copilot yielded an accuracy improvement of 275.15\%***, which means that Copilot subjects, on average, reported almost 3.7 times more relevant facts.

We also note that some subjects answered zero out of four MC questions correctly (eight in each group), which is worse than random guessing. Taking these subjects away, Copilot's accuracy effect on the device policy management MC questions becomes statistically significant at 11.39\%*. This is reported in table \ref{tab:accuracy} under ``MC $(Points\geq 1)$.''  

\begin{figure}[h]
    \centering
    \includegraphics[width=\linewidth]{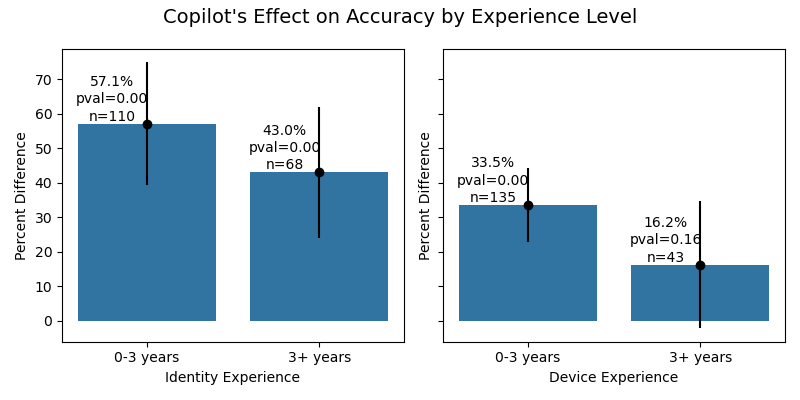}
    \caption{Copilot's effect heterogeneity by experience level and domain.}
    \label{fig:accexp}
\end{figure}

Consistent with prior findings, our point estimates suggest less experienced workers may benefit more from Copilot. However, we do not find that the differences in estimated effect sizes are statistically significant. This is illustrated in figure \ref{fig:accexp}, where the left panel splits the Copilot effect for sign-in troubleshooting by years of identity management experience. Both the less-experienced subjects (0-3 years) and the more-experienced subjects (3+ years) experience gains, 57.1\%*** and 43.0\%***, respectively. Although these estimates are statistically significantly different from zero, they are not statistically significantly different from each other. Note the 90\% confidence intervals (black bars) are overlapping heavily.\footnote{We also confirm this result in a regression framework with interaction terms to capture heterogeneous treatment effects and observe a lack of statistical significance on those terms.} In the right panel of figure \ref{fig:accexp}, we show effects by experience level for network device management. Once again, the order of the point estimates is consistent with Copilot conferring greater benefits for less-experienced subjects (33.5\%*** versus 16.2\%), but the difference in these point estimates is not statistically significant. 

Finally, we note that Copilot subjects' free responses scored 100.17\%*** higher in terms of their clarity and organization. As with accuracy, our point estimates suggest that Copilot offers greater benefits to less experienced subjects, but the differences are not statistically significant.

\subsection{Speed}
\label{sec:speed}
We measure statistically significant time savings in all scenarios. Table \ref{tab:speed} summarizes these results by task type and scenario. The ``Aggregate Results, All Content (Holding accuracy constant)'' numbers represent our headline speed improvements, a 29.79\%*** time savings for Copilot users. We also note that Copilot generally involves greater latency than typical page-loading in the admin centers, which is common among GAI tools. This necessarily slowed the Copilot users. Product improvements should reduce this latency and further increase the time savings for users with Copilot.

\begin{table}[ht]
\centering
\begin{tabular}{l c c c c}
\toprule
& \textbf{Overall} & \textbf{Treatment} & \textbf{Control} & \textbf{Improvement} \\ 
\midrule
\multicolumn{5}{l}{\textbf{Aggregate Results}} \\ 
All Content & 53.84 & 44.29 & 63.10 & -29.79\%*** \\ 
SATA & 17.81 & 15.25 & 20.14 & -24.28\%*** \\
MC & 10.91 & 12.02 & 9.60 & 25.13\%*** \\
MC \textit{(Drop first)} & 7.17 & 7.47 & 6.71 & 10.92\% \\
Free Response & 30.98 & 17.03 & 43.82 & -61.14\%*** \\ 
\midrule
\multicolumn{5}{l}{\textbf{Sign-In Troubleshooting}} \\ 
All Content & 24.04 & 16.63 & 30.46 & -45.41\%*** \\ 
SATA & 6.41 & 5.04 & 7.63 & -33.98\%*** \\ 
Free Response Content & 17.37 & 11.59 & 22.40 & -48.23\%*** \\ 
\midrule
\multicolumn{5}{l}{\textbf{Device Policy Management}} \\ 
All Content & 20.00 & 17.45 & 22.23 & -21.64\%*** \\ 
MC & 10.91 & 12.02 & 9.60 & 25.13\%*** \\
MC \textit{(Drop first)} & 7.17 & 7.47 & 6.71 & 10.92\% \\
Free Response Content & 12.17 & 5.43 & 19.27 & -71.80\%*** \\ 
\midrule
\multicolumn{5}{l}{\textbf{Device Troubleshooting}} \\ 
SATA & 11.24 & 10.21 & 12.11 & -15.69\%*** \\ 
\bottomrule
\end{tabular}
\caption{Comparison of task duration (minutes) by task type, holding accuracy constant\label{tab:speed}}
\label{tab:accuracy}
\end{table}

We measure time savings for Copilot across all three scenarios. The time savings included 45.41\%*** for sign-in troubleshooting, 21.64\%*** for device policy management, and 15.69\%*** for device troubleshooting. 

\begin{figure}[!ht]
    \centering
    \includegraphics[width=0.8\linewidth]{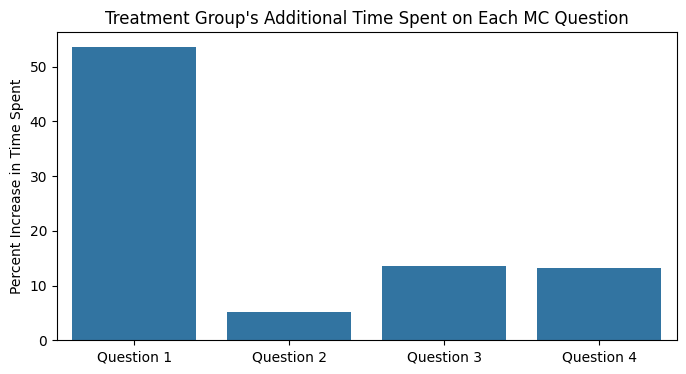}
    \caption{The difference in time spent between Copilot and control users is greatest on the first MC question, suggesting a possible learning phenomenon.\label{fig:MCtiming}}
\end{figure}

We find that Copilot's time savings track its accuracy improvements across task types. That is, the biggest gains again come from free response (61.14\%*** time savings), followed by SATA (24.28\%***). Lastly, the MC questions show a \textit{decrease} in speed for Copilot users (25.13\%*** time increase). For these MC questions we also estimate a negative relationship between accuracy and time spent for the control group. This is likely because control subjects who knew the answer responded quickly, and those who did not know it gained little accuracy by spending more time by, for example, searching the internet. This can explain the data we see, where even though Copilot provided an advantage in responding accurately, controlling for accuracy in calculating its time savings does not make up for the disadvantage of its latency. 

We also notice that Copilot users spent much more time than the control users on the first MC question compared to the subsequent three questions (see figure \ref{fig:MCtiming}). We think this suggests Copilot users were learning how to use Copilot for this type of task. Therefore, we drop the first question and recompute the time differences across the remaining tasks, which results in no statistically significant difference in task completion time (see the ``Drop first'' entry in table \ref{tab:speed}).  

\begin{figure}[ht]
    \centering
    \includegraphics[width=.9\linewidth]{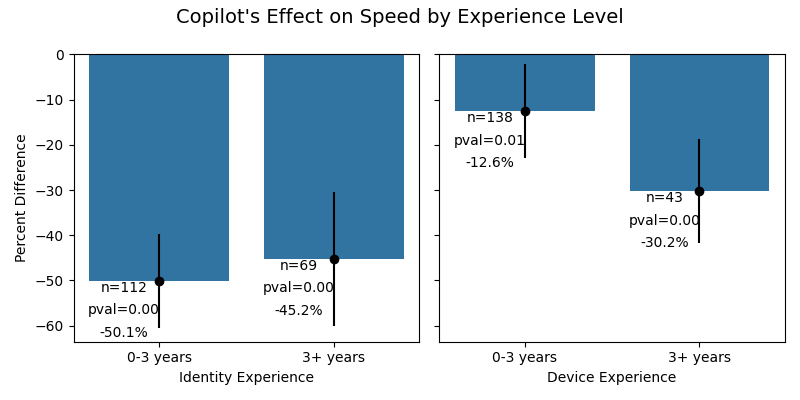}
    \caption{Copilot's effect heterogeneity by experience level on speed, holding accuracy constant}
    \label{fig:speedxp}
\end{figure}

As we did with accuracy, we examine time savings across different experience levels and domains. Our findings once again reveal statistically significant point estimates of time savings for all groups, as shown in figure \ref{fig:speedxp}. However, we do not observe statistically significant differences in estimated effect sizes \textit{between} any of the groups. The left panel illustrates Copilot's impact on task completion times for the sign-in troubleshooting task, with a 50.1\%*** and 45.2\%*** time savings for Copilot users with 0-3 and 3+ years of experience, respectively. These point estimates are consistent with intuition and our accuracy findings. 

The right panel of figure \ref{fig:speedxp} shows time savings of 12.6\%*** and 30.2\%*** for 0-3 and 3+ years of experience, respectively, for device management tasks. This suggests that experienced network management subjects benefit more from Copilot in terms of time savings than their less experienced counterparts, which runs counter to intuition and expectation. This is driven by the fact that less experienced control subjects are driving the negative relationship between task completion time and accuracy  for the device policy management MC question, mentioned above. This shrinks the possible time savings for the less experienced Copilot subjects. 

\subsection{Sentiment}
\label{sec:sentiment}
Subjects who used Copilot rated it favorably. We presented them with the following statements, with sliders to range from complete disagreement (scored as 0) to complete agreement (100). We report their responses in table \ref{tab:sentCopilotonly}. The ``Rating'' column reports the mean responses from the subjects. The lowest of these means is 87.51 for ``Copilot reduced my effort on this task." The highest is 96.77 for ``I would want to have Copilot the next time I do this task.'' The ``Percent Agreeing'' column reports the proportion of Copilot users with responses greater than 50. The lowest here is 96\% agreeing that Copilot reduced their effort. There was either unanimous or near-unanimous agreement that Copilot made subjects more productive, improved the quality of their work, and that they would want to have Copilot the next time they did these tasks.

\begin{table}[ht]
\centering
\begin{tabular}{l c c}
\toprule
\textbf{Statement} & \textbf{Rating} & \textbf{Percent Agreeing}\\ 
\midrule
Copilot reduced my effort on this task & 87.51 & 0.96\\ 
Copilot made me more productive & 92.84 & 0.99\\ 
Copilot helped me improve the quality of my work & 95.09 & 1.00 \\ 
I would want to have Copilot the next time I do this task & 96.77 & 1.00 \\ 
\bottomrule
\end{tabular}
\caption{Summary of Copilot-only sentiment responses. \label{tab:sentCopilotonly}}
\end{table}

We also asked both treatment and control users their agreement with standard statements about 
the task. Responses favored Copilot for seven of the nine statements, including the only two statements for which differences were statistically significant.  

\begin{table}[ht]
\centering
\begin{tabular}{l c c c}
\toprule
\textbf{Statement} & \textbf{Treatment} & \textbf{Control} & \textbf{Percent Change} \\ 
\midrule
\textbf{Positive} & & & \\
I felt effective doing this task & 88.28 & 87.90 & 0.43\% \\ 
I felt productive doing this task & 92.15 & 89.74 & 2.69\% \\ 
I felt in control while doing this task & 85.92 & 87.45 & -1.75\% \\ 
I felt secure while doing this task & 92.55 & 89.38 & 3.56\% \\ 
I would like a job like this as my full-time job & 89.11 & 86.83 & 2.62\% \\ 
\midrule
\textbf{Negative} & & & \\
I felt inadequate while doing this task & 21.30 & 19.41 & 9.72\% \\ 
I felt uncertain while doing this task & 21.71 & 24.34 & -10.82\% \\ 
This task was a lot of effort & 32.40 & 44.81 & -27.70**\% \\ 
This task was draining & 21.23 & 37.34 & -43.14***\% \\ 
\bottomrule
\end{tabular}
\caption{Summary of sentiment responses that applied to both groups. Positive statements are above the middle line, and negative statements are below it.\label{tab:sent}}
\end{table}

\begin{figure}[h]
    \centering
    \includegraphics[width=0.5\linewidth]{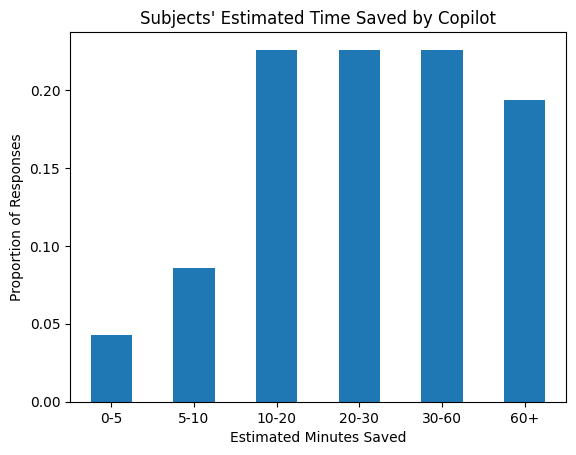}
    \caption{Subjects' responses when prompted to estimate how much time Copilot saved them. \label{fig:esttime}}
\end{figure}

As reported in table \ref{tab:speed} above, Copilot saved treatment subjects, on average, 18.23*** minutes (holding accuracy constant). However, when asked to estimate the savings, 64.5\% of subjects reported saving more than 20 minutes, and 41.9\% estimated it saved them more than 30 minutes. Subjects' overestimates of the time Copilot saved them, shown in figure \ref{fig:esttime}, are consistent with subjects enjoying using Copilot and appreciating they had access to it.

\section{Discussion}
The results of this study provide compelling evidence that GAI tools like Microsoft's Security Copilot can deliver substantial productivity improvements, in both accuracy and speed, for IT admins. These findings align with existing literature on the impact of GAI in other domains, like software development and customer support, reinforcing the idea that GAI tools are particularly effective in enhancing skilled workers' performance by automating routine tasks and providing insights more efficiently. Future iterations of Copilot will continue to improve latency, which could further amplify productivity gains.

However, where some studies have identified effect heterogeneity by skill and experience levels, these differences were not statistically significant in our study. That said, the combination of the evidence from other studies, our intuition about the potential gains by worker skill, and our consistent point estimates suggests that such heterogeneous effects may also apply to GAI in the IT admin domain. 

It is difficult to generalize about which task type best reflects IT admin work. This is, in part, why we included all three in our experiment. Our finding that Copilot's improvements are correlated with the task complexity suggests that IT admins will experience the greatest benefit from Copilot on scenarios that resemble our free response tasks, open-ended problem-solving requiring IT admins to synthesize information across multiple sources. This study also suggests that Copilot is likely to confer more modest, though still significant, improvements for scenarios that align to SATA and MC tasks, i.e., where the challenge is to choose from a limited set of predefined solutions. Although we report average effects across all scenarios and task types, the more detailed breakdown of effects by scenario and task type are a better guide to the gains that practitioners can expect to realize. 

In terms of deployment, organizations should carefully consider where GAI tools will have the most impact. This study indicates that GAI’s greatest utility for IT admins is in open-ended tasks requiring synthesis of multiple data points, so focusing on those areas may yield the most significant efficiency gains. Nevertheless, integrating GAI into simpler tasks should not be dismissed outright, as even small reductions in cognitive load can contribute to overall job satisfaction and performance over time. 

The results from laboratory experiments provide powerful evidence about GAI's productivity effects, but no laboratory can exactly replicate the nuances of live operations. Questions remain about how much of the observed effects will translate to live operations. Hence, we look forward to future work utilizing field experiments involving IT admins to fully cement the existence of the effects measured here.

\paragraph{Acknowledgements} 
We would like to thank the following individuals for their critical contributions to the set up and execution of this experiment: A-Shawni Mitchell, Ashley Torres Ventura, Ari Schorr, Chung-Wei Foong, Diana Vicezar, Evan Greenberg, Janelle Bryant, Justin Grana, Kleanthis Karakolios, Katerina Marazopoulou, Lavanya Lakshman, Mitch Muro, Nandini Bhatt, Sarah Scott, Sean Cahill, Shravana Mukherjee, Ankit Shrivastava, and Tracy Shi.

\bibliography{refs}

\appendix
\section{Instruction Clarity}
\begin{figure}[!ht]
    \centering
    \includegraphics[width=0.65\linewidth]{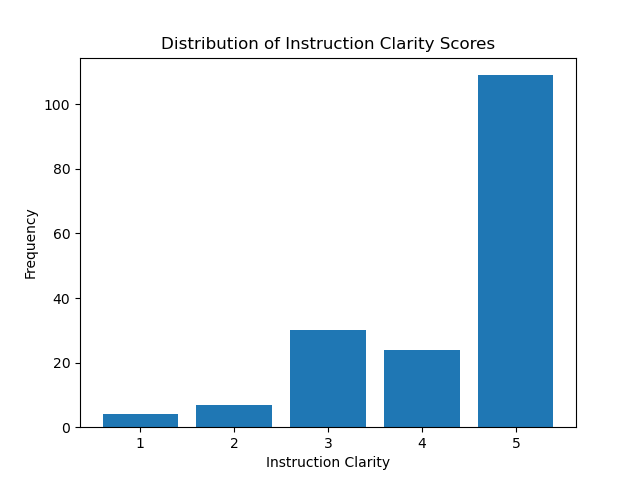}
    \caption{Subjects' responses when prompted to provide instruction clarity feedback \label{fig:instructclarity}}
\end{figure}

Participants were asked to evaluate the clarity of the provided instructions. Their responses were rated on a scale from 1 to 5, where 1 indicated confusion and lack of clarity, and 5 indicated clear and actionable instructions. Figure \ref{fig:instructclarity} presents a density plot illustrating the distribution of these ratings. The results show high scores for instruction clarity, with only 11 participants rating the instructions as 1 or 2. Notably, 62.64\% of participants reported no issues or confusion with the provided instructions.

\section{Calculating Time Savings Holding Accuracy Constant}

Note that a simple regression of time on treatment and score assumes the returns to additional time spent are the same between the treatment and control groups, which is unlikely true. Therefore, to compare speed holding accuracy constant, we employ a linear regression framework that proceeds in three steps. First, we estimate the task duration as a function of accuracy for the control group. Then we predict the task duration the control group would need to achieve the same accuracy as the Copilot group. Finally, we compute the difference between task durations for the two groups, using a bootstrap to compute the level of certainty of this finding.

Let $T_g$ and $A_g$ represent the set of task durations and accuracy scores for each subject in group $g\in\{(t)reatment,(c)ontrol\}$. And let $\bar{T}_g$ and $\bar{A}_g$ represent the sample means. We then estimate the effect of accuracy on task duration for the control group via the following regression and ordinary least squares. 
$$T_c = \alpha +\beta A_c + \epsilon$$
With $\hat{\alpha}$ and $\hat{\beta}$, we next solve for the time it would take the control group to achieve $\bar{A}_t$, and call this $\Tilde{T}_c \equiv \hat{\alpha}+\hat{\beta}\bar{A}_t$. Let the difference $\Tilde{T}_c - \bar{T}_t$ be the effect of Copilot on task duration holding accuracy constant. Then, to perform inference, we simply bootstrap this procedure by sampling with replacement the control subjects and calculating $\Tilde{T}_c - \bar{T}_t$ for each sample to get our bootstrap distribution of the effect of Copilot on task duration holding accuracy constant.
\end{document}